\documentclass[floatfix,showpacs,preprintnumbers,amsmath,amssymb,aps,twocolumn,superscriptaddress,pra,10pt]{revtex4-1}

\usepackage{amsfonts}
\usepackage[pdftex]{graphicx}
\usepackage{color}
\usepackage{bm}
\usepackage{multirow}
\usepackage{comment} 

\newcommand{\figref}[1]{Fig.~\ref{#1}}
\newcommand{\e}[1]{\text{e}^{#1}}
\newcommand{\cmplxi}{\text{i}}
\newcommand{\diffd}{\text{d}}
\newcommand{\tr}{\operatorname{Tr}}
\renewcommand{\vec}[1]{\mathbf{#1}}
\newcommand{\punc}[1]{\,#1}
\newcommand{\neweqnline}{\nonumber\\}

\newcommand{\eqnref}[1]{Eqn.~(\ref{#1})}
\newcommand{\vecgrk}[1]{\boldsymbol{#1}}

\begin{document}

\title{Quantum Monte Carlo study of the two-dimensional ferromagnet}
\author{G.J.~Conduit}
\affiliation{Cavendish Laboratory, J.J. Thomson
  Avenue, Cambridge, CB3 0HE, United Kingdom}
\date{\today}

\begin{abstract}
  We present Quantum Monte Carlo calculations that probe the
  paramagnet-ferromagnet phase transition in a two-dimensional Stoner
  Hamiltonian. With a screened Coulomb interaction we observe a first order
  ferromagnetic transition for short screening lengths, and a second order
  transition with a longer screening length, accompanied by a rising
  critical interaction strength. Finally, we discuss the consequences of our
  results for an ultracold atomic gas with finite ranged interactions.
\end{abstract}


\maketitle

\section{Introduction}

Layered systems that display magnetic correlations have emerged as an
important testbed of strongly correlated physics. The Stoner Hamiltonian
represents the simplest possible metallic system that undergoes a
ferromagnetic transition. Since a mean-field analysis~\cite{Conduit2D}
predicts a second order transition, in the vicinity of the low temperature
transition Hertz~\cite{Hertz1976} predicted that quantum fluctuations would
drive critical behavior. However, it has recently been predicted that
quantum fluctuations are even stronger than envisaged by Hertz and drive
phase reconstruction through a first order ferromagnetic
transition~\cite{Conduit08,Conduit09,Conduit2D,Belitz05,Maslov06,Efremov08,Karahasanovic12},
consistent with the phase transition observed in the quasi two-dimensional
systems Sr$_{3-x}$Ca$_{x}$Ru$_{2}$O$_{7}$~\cite{Ikeda97,Perry04},
Sr$_{2}$RuO$_{4}$~\cite{Mazin97},
La$_{x}$Sr$_{2-x}$RuO$_{4}$~\cite{Kikugawa04},
Ca$_{2}$RuO$_{4}$~\cite{Nakamura02}, and
UGe$_{2}$~\cite{Saxena00,Huxley01,Watanabe02}. The importance of the quantum
fluctuations in driving this phase reconstruction motivates a careful
theoretical analysis of the Stoner Hamiltonian.  However, analytical
studies~\cite{Conduit08,Conduit09,Conduit2D,Belitz05,Maslov06,Efremov08,Karahasanovic12}
of the magnetic transition rely on a perturbation theory in the interaction
strength, whereas in reality the interactions are strong and the quantum
fluctuations dominant. To probe the phase transition in the non-perturbative
regime we perform and present the first Quantum Monte Carlo (QMC)
calculations of the two-dimensional itinerant ferromagnet with short ranged
interactions. We complement this with the first analytical study of the
ferromagnetic transition in two dimensions with a screened Coulomb
interaction that allows us to be the first to calculate the dependence of
the tricritical point temperature on screening length.  The QMC calculations
employ a fixed node approximation, tempered by backflow corrections, so
should complement and improve upon the accuracy of the analytical findings.

An ultracold atomic gas could be an attractive alternative realization of
the Stoner Hamiltonian. Recent experiments on the three-dimensional
system~\cite{Jo09}, and polaron systems~\cite{Schirotzek09,Kohstall12} have
delivered some evidence for ferromagnetic
ordering~\cite{Duine05,LeBlanc09,Conduit09ii}, though a competing loss
process provides an alternative
explanation~\cite{Pekker11,Sanner12}. Accurate QMC calculations that pin the
transition down should help guide future experiments that could realize a
Hamiltonian analogous to the Stoner
system~\cite{Conduit08,Conduit10,Conduit2D,vonKeyserlingk11,Baur12,Bugnion13i,Bugnion13ii}. The
theoretical and experimental study of the idealized two-dimensional
ultracold atomic gas also presents an opportunity to shed light on high
temperature superconductivity where antiferromagnetism competes with the
d-wave superconducting phase~\cite{Conduit2D}.

\begin{figure}
 \includegraphics[width=1.\linewidth]{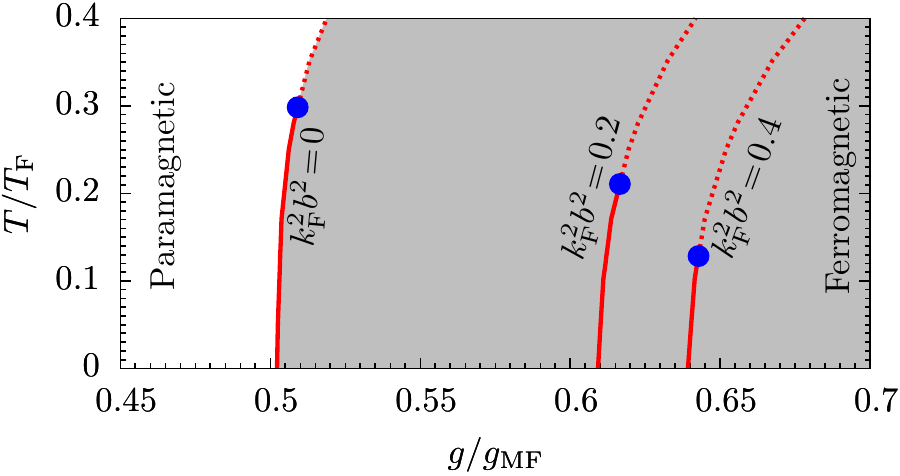}
 \caption{(Color online) The phase diagram for the system from the
  paramagnetic to ferromagnetic phase (shaded gray). The red solid
  line shows the first order ferromagnetic transition and the dashed
  line the second order transition with the tricritical point
  highlighted by the blue dot. The three lines denote boundaries
  with a screening length of $k_{\text{F}}^2b^2=0$,
  $k_{\text{F}}^2b^2=0.2$, and $k_{\text{F}}^2b^2=0.4$.}
 \label{fig:PhaseDiagram}
\end{figure}

In this paper we present QMC calculations to analyze the
paramagnet-ferromagnet transition in two dimensions. To connect to the solid
state the QMC calculations are performed first with the screened Coulomb
inter-particle potential and later the results are compared to the square
well potential. Both inter-particle potentials are characterized by a range
parameter that we vary to gauge the consequences of screening in the solid
state, and the interaction effective range in the cold atom gas.  Combining
our QMC results with a complementary analytical order-by-disorder approach
allows us to derive the phase diagrams shown in \figref{fig:PhaseDiagram} at
finite temperatures.  The phase diagram shows that the
paramagnet-ferromagnet transition reverts from first to second order on
increasing the screening length, and the corresponding critical interaction
strength increases. The introduction of a finite interaction range with
$k_{\text{F}}b>0$ increases the critical interaction strength, and lowers
the tricritical point temperature to be in line with experimental values.
Finally, we adapt our formalism to assess the opportunity to observe
ferromagnetic correlations in an ultracold atomic gas.

\section{Formalism}

To model the ferromagnet we focus on the modified Stoner Hamiltonian
\begin{align}
 \hat{H}=\sum_{\vec{p},\sigma}
 \epsilon_{\bf p}\hat n_{{\bf p}\sigma}
 +\iint\diffd\vec{r}_{\uparrow}\diffd\vec{r}_{\downarrow}
 g(\vec{r}_{\uparrow}-\vec{r}_{\downarrow})
 \hat n_{\vec{r}_{\uparrow}\uparrow}
 \hat n_{\vec{r}_{\downarrow}\downarrow}\punc{,}
 \label{Hamiltonian}
\end{align}
where $\epsilon_{\vec{p}}=p^2/2$ is the dispersion, $n_{\vec{p}\sigma}$ is
the fermion occupancy in momentum space, $n_{\vec{r}\sigma}$ in real space,
spin $\sigma\in\{\uparrow,\downarrow\}$, and we adopt atomic units with
$\hbar=m=1$ throughout.  To simulate the inter-particle repulsion we use a
screened Coulomb interaction parameterized as $g(r)=g\e{-r/b}/2\pi br$
acting between opposite spin particles separated by a distance $r$. It is
characterized by a screening radius $b$ and interaction strength $g$ To
validate our results we also study the square potential of radius $R$,
$g(r)=g\theta(R-r)/\pi R^2$ with Heaviside function $\theta$, whose
interaction strength and screening parameter are related to the screened
Coulomb interaction through a momentum-space expansion~\cite{Thankappan93}.
The interparticle potentials are defined so that they have the simple
momentum space forms $g_{\vec{q}}=g/\sqrt{1+b^2 q^2}$ and $g(1-R^2q^2/8)$
respectively. This allows them to be used within our analytical
formalism. These definitions mean that the ferromagnetic transition in the
mean-field approximation emerges at the critical interaction strength
$g_{\text{MF}}=2\pi$.  To evaluate the energy of the electrons we employ two
complementary techniques, a Quantum Monte Carlo calculation limited only by
a fixed node approximation, and an analytical evaluation of the free energy
derived by a functional integral formalism. The mechanics of both methods
are outlined below before we study the resulting phase diagrams.

\subsection{Quantum Monte Carlo formalism}

\begin{figure}
 \includegraphics[width=1.\linewidth]{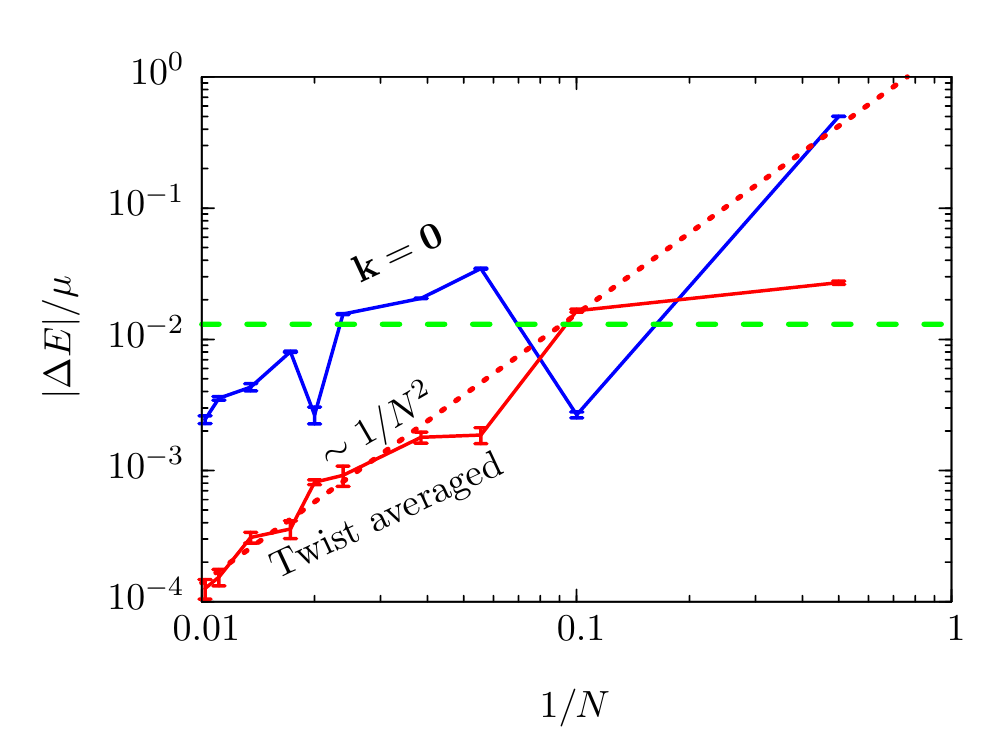}
 \caption{(Color online) The error in energy $\Delta E$ for the
   $\vec{k}=\vec{0}$ (blue) and twist averaged (red) calculations with
   system size ($N$) for the screened Coulomb potential with $b=0$,
   $g/g_{\text{MF}}=0.5$. The red dotted line shows the $\sim1/N^2$
   scaling. The green dashed line shows the energy resolution required to
   determine the order of the ferromagnetic transition.}
 \label{fig:LengthExtrapolationPlot}
\end{figure}

To seek and calculate the ground state of the Hamiltonian we perform QMC
simulations with the code \textsc{casino}~\cite{Needs10}. This method
optimizes a trial wave function at zero temperature, and finds the exact
ground state subject to the nodal surface of the wave function being
fixed. The approach is a refinement of that used in previous studies of
itinerant
ferromagnetism~\cite{Ceperley80,Ortiz99,Zong02,Conduit09,Pilati10,Chang11,Drummond11}. We
use a variational wavefunction $\psi=\e{-J}D$ that is a product of a Slater
determinant, $D$, which takes full account of the fermion statistics and
includes further electron-electron correlations through a Jastrow factor
$J$. The QMC simulations comprise of two stages: firstly in variational
Monte Carlo (VMC) the ground state energy was minimized by varying the
parameters in the Jastrow factor; secondly diffusion Monte Carlo (DMC)
starts from the VMC wave function, and treats the Schr\"odinger equation as
a diffusion equation to project out the exact ground state subject to a
fixed node approximation.

The Slater determinant $D=\det(\{\psi_{\vec{k}\in
  k_{\text{F}\uparrow}},\bar{\psi}_{\vec{k}\in k_{\text{F}\downarrow}}\})$
consists of plane-wave orbitals
$\psi_{\vec{k}}(\vec{r})=\exp(\cmplxi\vec{k}\cdot\vec{r})$ whose momenta
$\vec{k}$ satisfy periodic boundary conditions in the square simulation
cell, and lie within the up/down spin Fermi surfaces $k_{\text{F}\sigma}$.
With a square simulation cell, to ensure that within VMC we have a
real-valued wave function and that the Fermi surface is circular, the number
of states must correspond to closed shells containing
$N_{\sigma}=\{1,5,9,13,21,25,29,37,45,49\}$ electrons, thus constraining us
to discrete values of the magnetization. For computational efficiency we
factorize the Slater determinant into up and down-spin
components~\cite{Needs08}, so $D=\det(\{\psi_{\vec{k}\in
  k_{\text{F}\uparrow}}\})\det(\{\bar{\psi}_{\vec{k}\in
  k_{\text{F}\downarrow}}\})$. Provided that the orbitals of the minority
spin state are the lowest energy orbitals of those in the majority spin
state,~\cite{Roothaan60}, this is the state with total spin
$s=s_{\text{z}}=(N_{\uparrow}-N_{\downarrow})/2$.

The Jastrow factor, $J(\{\vec{r}_{i}\})$, accounts for electron-electron
correlations. It has the general form
\begin{align}
 J(\{\vec{r}_{i}\})=\sum_{i=1}^{N-1}\sum_{j=i+1}^{N}u(r_{ij})+p(\vec{r}_{ij})\punc{,}
\end{align}
where the summation over indices $\{i,j\}$ cover all
$N=N_{\uparrow}+N_{\downarrow}$ electrons, the
electron separation is $\vec{r}_{ij}=\vec{r}_{i}-\vec{r}_{j}$, and
$r_{ij}=|\vec{r}_{ij}|$. The Jastrow factor includes the polynomial
expansion in electron-electron separation proposed in Ref.~\cite{Drummond04}
\begin{align}
 &u(r_{ij})=(L-r_{ij})^{3}\Theta(L-r_{ij})\times\neweqnline
&\left[\alpha_{0}+r_{ij}
\left(\frac{3\alpha_{0}}{L}-\frac{\Gamma_{ij}}{L^3}\right)+
\sum_{l=2}^{N_{\text{u}}}\alpha_{l}r_{ij}^{l}\right]\punc{,}
\end{align}
chosen so that it satisfies the Kato cusp conditions at $r_{ij}=0$,
is zero beyond the cutoff length $L$ imposed by the Heaviside function
$\Theta$, $\Gamma_{ij}=1/4$ for equal spin electrons and $\Gamma_{ij}=1/2$
for opposite spins, and contains $N_{\text{u}}=8$ variational parameters
$\{\alpha_{l}\}$. The cutoff length $L$ was chosen to be the largest circle
that could be inscribed in the simulation cell. The Jastrow factor also
includes a plane-wave expansion
\begin{align}
 &p(\vec{r}_{ij})=\sum_{A=1}^{N_{\text{p}}}a_{A}
\sum_{\vec{G}_{A}^{+}}\cos(\vec{G}_{A}\cdot\vec{r}_{ij})\punc{,}
\end{align}
where the $\{\vec{G}_{A}\}$ are the reciprocal lattice vectors of the
simulation cell belonging to the $A$th shell of vectors under the full symmetry group
of the Bravais lattice, and the superscript ``$+$'' means that if $+\vec{G}_{A}$ is
included in the sum then $-\vec{G}_{A}$ is excluded. The summation over $A$
covers $N_{\text{p}}=8$ shells, with corresponding variational parameters
$\{a_{A}\}$. To broaden the freedom of the variational wave function we also
include backflow corrections~\cite{LopezRios06}. These substitute the
electron coordinates $\vec{r}_{i}$ in the Slater determinant a new set of
collective coordinates
$\vec{x}_{i}=\vec{r}_{i}+\boldsymbol{\xi}_{i}(\{\vec{r}_{j}\})$ where the
backflow displacement of electron $i$ is $\boldsymbol{\xi}_{i}$ given by
\begin{align}
 \boldsymbol{\xi}_{i}=\sum_{j\ne i}^{N}\vec{r}_{ij}\left(1-\frac{r_{ij}}{L}\right)^3
\Theta(L-r_{ij})
\sum_{k=0}^{N_{\eta}}c_{k}r_{ij}^{k}\punc{,}
\end{align}
where $L$ is the cutoff length, $N_{\eta}=7$ the expansion order, and
$\{c_{k}\}$ the variational parameters. The inclusion of the backflow
corrections allows the nodal surface of the wave function to shift and
therefore relax the fixed node approximation. To seek the ground state the
variational parameters of the trial wave function were numerically optimized
within VMC by minimizing the total VMC energy~\cite{Umrigar07}. The
optimized VMC wave function was used as the trial state for the DMC
calculation.

The DMC method~\cite{Needs08} simulates a population of walkers whose
evolution is driven by the imaginary time Schrodinger equation to project
out the ground state component of the VMC wave function. The walk is taken
in a series of discrete time steps, with walkers branching or annihilating
according to the local energy. The choice of time step, the control of the
walker population, and the system size can each introduce errors into the
final prediction of the ground state energy. We therefore now address how to
minimize each source of uncertainty in turn.

\emph{Time step}: One can propagate forwards in time exactly by using
Green's function Monte
Carlo~\cite{Kalos62,Kalos67,Kalos74,Ceperley81,Reynolds82,
  Ceperley84,Ceperley86,Schmidt87} but unfortunately the method is
computationally expensive~\cite{Foulkes01,Needs10}. Therefore here we employ
an approximate Green's function that would become exact in the limit of
short time-steps $\tau$ for the walker evolution. However, the computational
effort required to achieve a given uncertainty in the prediction for the
ground state energy increases as $1/\tau$, ruling out the use of
infinitesimal time steps in practice. Therefore, where high accuracy is
required, we use two different finite time steps $\{\tau_{i}\}$ and
extrapolate to $\tau=0$ to obtain the ground-state energy. Tests revealed
that the time step error had entered the linear regime at $\tau<0.01$ (at a
Wigner-Seitz radius $r_{\text{s}}=1$). Here to minimize the uncertainty in
the extrapolate we followed the prescription of Lee~\cite{Lee11} and ran
simulations with time steps $\tau=0.01$ and $\tau/4=0.0025$ for relative
durations of $1:8$ respectively, and then finally extrapolated to $\tau=0$.

\emph{Population control}: To ensure that the population control bias is
negligible in all runs the target population exceeded
2000~\cite{Drummond08}.  The number of equilibration steps discarded during
the equilibration phase of each DMC calculation was set so that the
root-mean-square distance diffused by a typical electron exceeded the
simulation cell size.

\emph{System size}: The VMC and DMC simulations must be performed in a
finite sized simulation cell that is periodically repeated to create an
effectively infinite system. However, using the finite sized simulation cell
introduces errors into the final prediction for the energy. The error in the
estimated energy due to the finite sized simulation cell can be divided into
three components: single particle kinetic energy, Hartree energy, and
exchange-correlation energy. The error in the kinetic energy arises because
of the approximation of the circular Fermi surface by the discrete set of
k-vectors of the closed shells within the simulation cell. As the system is
enlarged the k-vector grid becomes more fine resulting in abrupt changes in
the kinetic energy. The use of a non-zero simulation cell Bloch wave vector
$\vec{k}_{\text{s}}$ that causes some k-vectors to lie outside of the Fermi
surface, and others within can lead to a dramatic reduction in the finite
size error, with the optimal being the Baldereschi
point~\cite{Baldereschi73}. However, it is even better to take an average
expectation over all Bloch vectors $\vec{k}_{\text{s}}$ within the first
Brillouin zone~\cite{Drummond08}. We adopted the most optimal strategy which
is to average over a uniform grid of twists (we use $1000$ points) centered
on the Baldereschi point of the simulation cell Brillouin
zone~\cite{Drummond08}. The Hartree energy is negligible due to the
short-ranged nature of the interaction whose maximum exponential decay
length is $~1/20$ of the typical simulation cell size. The final
contribution to the finite sized error is due to long-range two-body
corrections. The change in kinetic energy can be encapsulated by the
Chiesa-Holzmann-Martin-Ceperley approximation~\cite{Chiesa06}. It has been
shown that in two dimensions the scaling with long ranged interactions is
$\sim N^{-5/4}$ whereas the with short ranged interaction such as our
screened Coulomb interaction the scaling is expected to be $\sim
N^{-2}$~\cite{Drummond08,Drummond11}. We also note that it has been found
that other expectation values such as magnetic susceptibility can suffer
more from finite size effects, but here we focus only on the ground state
energy~\cite{Holzmann09}.

To gauge the scale of the finite size error we ran tests on a paramagnetic
system with the screened Coulomb interaction with $b=0$ and
$g/g_{\text{MF}}=0.5$ and vary the system size by changing the number of
electrons $N$. \figref{fig:LengthExtrapolationPlot} shows that twist
averaging delivers an error in the energy (determined against a $1/N^2$
extrapolation to the infinite sized system at $1/N=0$) that is over an order
of magnitude smaller than that from taking just the single result calculated
at $\vec{k}=\vec{0}$. The scaling $~1/N^2$ appears similar to that found in
a previous two-dimensional study with short-ranged
interactions.~\cite{Drummond11} For the typical system size used $2/N<0.02$
the finite sized error is almost two orders of magnitudes smaller than the
smallest energy scale $\Delta E\approx0.012$ that we will need to resolve
the features of the phase transition (see
\figref{fig:EnergyDiagram}(a,b)). For large magnetization the energy
associated with the minority spin species will have a finite size correction
that scales as $1/N_{\text{minority}}^2$ but since that species now makes
only a small contribution to the overall energy it is beyond the order
needed.  Therefore twist averaging ensures that finite sized errors are
inconsequential when analyzing the phase diagram. Finally we note that all
of our twist-averaged energies are always slight over estimates of the true
energy since the single-particle kinetic energy $k^2/2$ is a convex function
and the occupied k-space is a convex polyhedron~\cite{Drummond08}.

\emph{Changing polarization}: Previous studies of $^{3}$He have established
that Slater-Jastrow wave functions overestimate the unpolarized state
energy, thus favoring the ferromagnetic state. Including many-body backflow
corrections can help reduce the energy of the unpolarized state to better
align with experiment~\cite{Zong03,Holzmann06}. Here we include two-atom
backflow corrections, that for short ranged interactions have previously
been found to have a relatively small impact in the ground state energy
versus that for long-ranged interactions~\cite{Drummond09,Drummond11},
indicating that the QMC bias towards unpolarized states is less for shorter
ranged interactions. However, we also note that an alternative calculation
on a lattice with a smaller fixed node bias indicates that the ferromagnetic
transition is infinite order rather than first order.~\cite{Carleo11}

\subsection{Analytical formalism}

The itinerant ferromagnet has previously been analyzed using a functional
integral formalism. This will provide a useful complementary tool to study
the ferromagnetic transition. The formalism  calculates the quantum
partition function from a coherent state field integral
\begin{align}
&\mathcal{Z}=\!\int\mathcal{D}\psi\text{exp}\Bigg[-\!\!\!\!
\sum_{p,\sigma=\left\{\uparrow,\downarrow\right\}}\!\!\!\!
\overline{\psi}_{p,\sigma}(-\cmplxi\omega+\xi_{\vec{p}\sigma})\psi_{p,\sigma}\neweqnline
&-\!\!\!\!\sum_{p_{\uparrow},p_{\downarrow},q}
\!\!\!\!g(\vec{p}_{\uparrow}\!-\!\vec{p}_{\downarrow})
\overline{\psi}_{\!p_{\uparrow}\!-\!q\!/\!2,\uparrow}\overline{\psi}_{\!p_{\downarrow}\!+\!q\!/\!2,\downarrow}
\psi_{\!p_{\downarrow}\!-\!q\!/\!2,\downarrow}\psi_{\!p_{\uparrow}\!+\!q\!/\!2,\uparrow}\!\Bigg]\!\punc{,}
\end{align}
where the field $\psi$ describes a two component Fermi gas. We now decouple
the quartic interaction term with a Hubbard-Stratonovich transformation into
the full vector magnetization $\vecgrk{\phi}$ and the density channel
$\rho$~\cite{Conduit08}. This is the simplest decoupling scheme that
maintains rotational spin invariance and yields the correct Hartree-Fock
equations~\cite{Hubbard79,Prange79}. With the action now quadratic in the
Fermionic degrees of freedom we integrate them out to recover the quantum
partition function
$\mathcal{Z}=\int\mathcal{D}\vecgrk{\phi}\mathcal{D}\rho\exp(-S)$ with the
action given by
\begin{align*}
  S\!=\!\tr\left[ \vecgrk{\phi} \hat{g} \vecgrk{\phi} \!-\!\rho\hat{g}\rho\right]\!
 -\!\tr\ln\!\left[(\hat{\partial}_{\tau}\!+\!\hat{p}^2/2\!-\!\mu\!
 +\!\hat{g}\rho)\mathbf{I}
 \!-\!\hat{g}\vecgrk{\phi}\!\cdot\!\vecgrk{\sigma}\right]\!\punc{.}
\end{align*}
Here we have employed the operator form of the interparticle potential
$\hat{g}$. We adopt $\vec{M}$ and $\rho_{0}$ as the putative saddle point
values of the fields $\vecgrk{\phi}$ and $\rho$, and then expand to
quadratic order in the fluctuations in those fields about the assumed saddle
point.  This allows us to integrate out the fluctuations in the
magnetization and density channels whilst keeping terms up to second order
in the interaction
strength~\cite{Conduit08,Conduit09,vonKeyserlingk13}. This yields the free
energy~\cite{Conduit08}
\begin{align}
&F=\sum_{\vec{p},\sigma}\frac{p^{2}}{2}
n_{\vec{p}\sigma}
+\!\sum_{\vec{p}_{\uparrow},\vec{p}_{\downarrow}}
g(0)
n_{\vec{p_{\uparrow}}\uparrow}
n_{\vec{p_{\downarrow}}\downarrow}\neweqnline
&+\!\!\!
\sum_{
 \renewcommand*{\arraystretch}{0.6}
 \begin{array}{c}
  \scriptscriptstyle\vec{p}_{\!1}+\vec{p}_{\!2}\\
  \scriptscriptstyle=\vec{p}_{\!3}+\vec{p}_{\!4}
 \end{array}
}
\!\!g^2(\vec{p}_{1}-\vec{p}_{3})
\frac
{n_{\vec{p_{1}}\uparrow}n_{\vec{p_{2}}\downarrow}
(n_{\vec{p_{3}}\uparrow}+n_{\vec{p_{4}}\downarrow})}
{\xi_{\vec{p}_{1}\uparrow}+\xi_{\vec{p}_{2}\downarrow}
-\xi_{\vec{p}_{3}\uparrow}-\xi_{\vec{p}_{4}\downarrow}}
\punc{,}
\label{eqn:FreeEnergy}
\end{align}
where the first term corresponds to the kinetic energy (with a Fermi
distribution $n_{\vec{p}\sigma}=1/(1+\e{\xi_{\vec{p}\sigma}/T})$,
$\xi_{\vec{p}\sigma}=p^2/2-\mu-\sigma gM$, and $T$ is the temperature), the
second is the mean-field contribution of the interactions, and the third
higher order interaction effects. The first two terms would be delivered by
the standard Stoner mean-field theory, and the final term is attributed to
fluctuation corrections~\cite{Conduit08}. We have opted, without loss of
generality, to set the quantization axis along the direction of the
magnetization. This expression for the free energy remains a function of the
magnetization and density, that can now be determined by minimizing the free
energy with respect to $M$ and $\rho$, thereby fulfilling our premise that
these are the saddle point values. Since the Fermi distributions have a
temperature dependence we can use the formalism to not only study $T=0$, but
also the phase behavior at finite temperature. The formalism applies not
only at zero temperature, but also at finite temperature, thus allowing us
to map out the entire phase diagram.

The screened Coulomb interaction in momentum space is
$g(p)=g/\sqrt{1+b^2p^2}$. We have included the effect of finite ranged
interactions into the free energy following the prescription in
Ref.~\cite{vonKeyserlingk13}, where it was applied to a three-dimensional
system. The mean-field term interaction strength is independent of momentum
exchange so is left unaffected by the screening length $b$. In the
fluctuation correction term the denominator means that dominant
contributions to the momentum summation arise at
$|\vec{p}_{1}-\vec{p}_{3}|=\sqrt{2}k_{\text{F}}$~\cite{Conduit09,vonKeyserlingk13}
so that we can simply adopt this fixed value within the interaction
$g(\vec{p}_{1}-\vec{p}_{3})\mapsto g(\sqrt{2}k_{\text{F}})$, and therefore
in the presence of screening the fluctuation correction term is simply
rescaled by a factor of
$1/(1+2k_{\text{F}}^2b^2)^2$~\cite{Conduit09,vonKeyserlingk13}. Here we use
the definition $k_{\text{F}}=\sqrt[3]{3\pi^2(n_{\uparrow}+n_{\downarrow})}$
where $n_{\sigma}$ is the density of the electrons with spin
$\sigma$. Following this rescaling, the ground state magnetization can again
be extracted by minimizing the free energy with respect to magnetization.

\section{Results}

\begin{figure}
 \includegraphics[width=1.\linewidth]{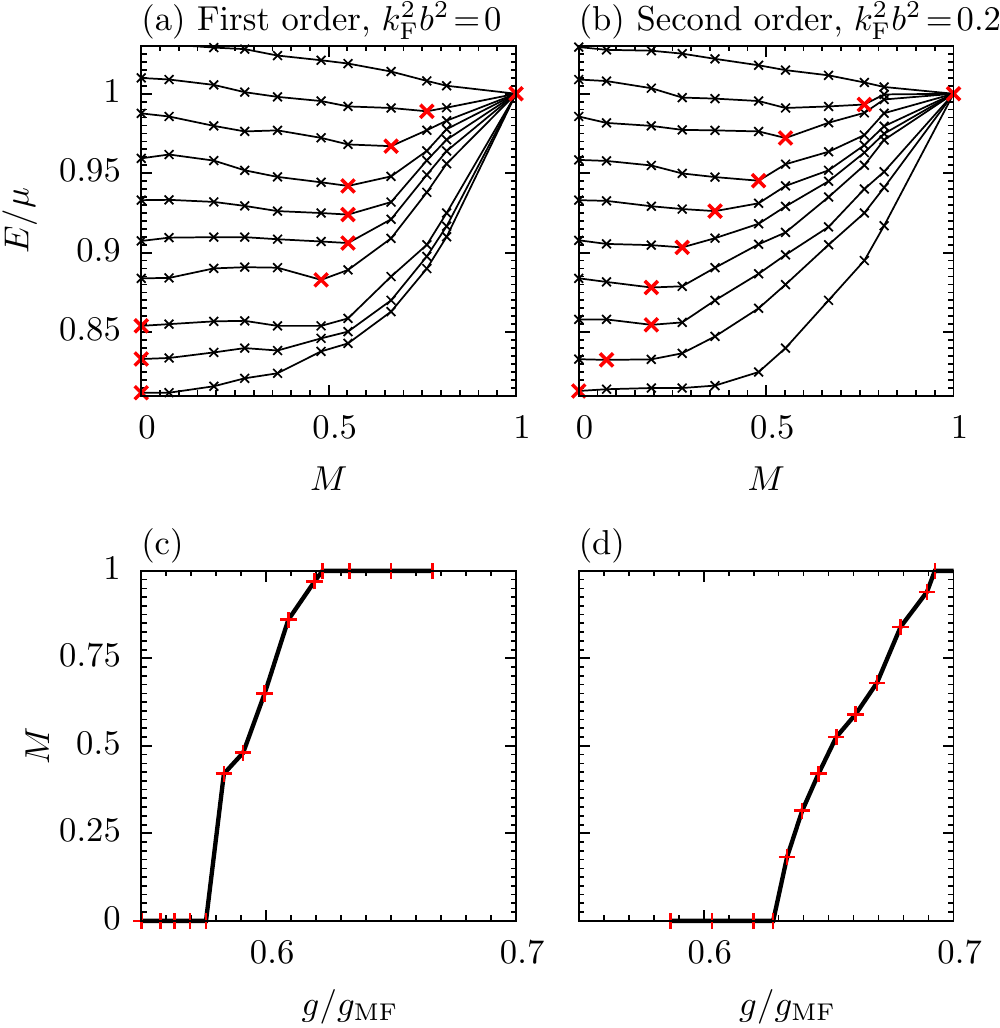}
 \caption{(Color online) (a,b) Energy bands as a function of magnetization
   $M$ at the interaction strengths shown in (c,d) for the screened Coulomb
   potential (the lowest band is the weakest interaction strength). The red
   points highlight the minimum point in each energy band. The error bars
   are approximately the line width. (c,d) Magnetization as a function of
   interaction strength across the transition. The left hand plots are at
   $k_{\text{F}}^2b^2=0$, and the right-hand at $k_{\text{F}}^2b^2=0.2$.}
 \label{fig:EnergyDiagram}
\end{figure}

\begin{figure}
 \begin{flushleft}(a) Zero temperature phase behavior\end{flushleft}
 \includegraphics[width=0.85\linewidth]{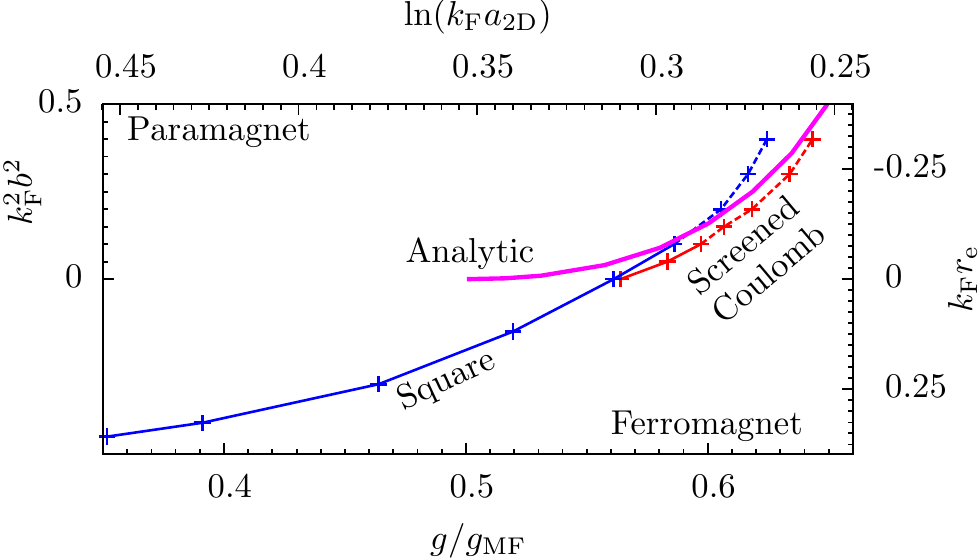}
 \begin{flushleft}(b) Zero temperature first order transition\end{flushleft}
 \includegraphics[width=0.7\linewidth]{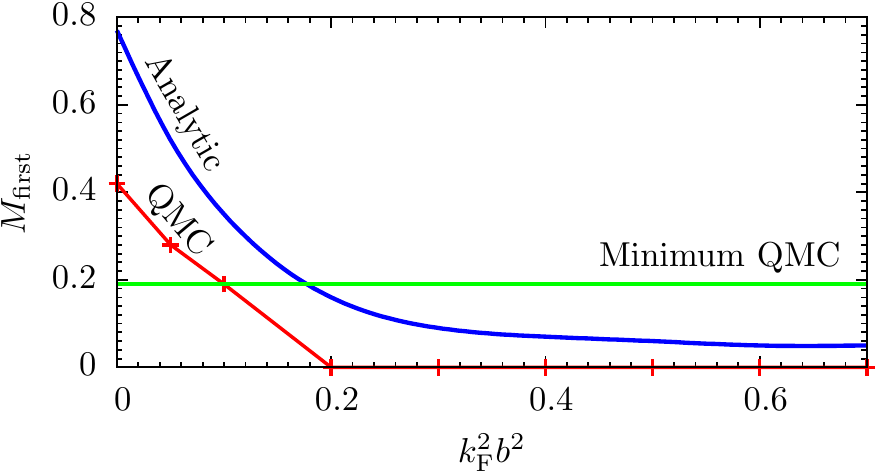}
 \begin{flushleft}(c) Finite temperature phase behavior\end{flushleft}
 \includegraphics[width=0.7\linewidth]{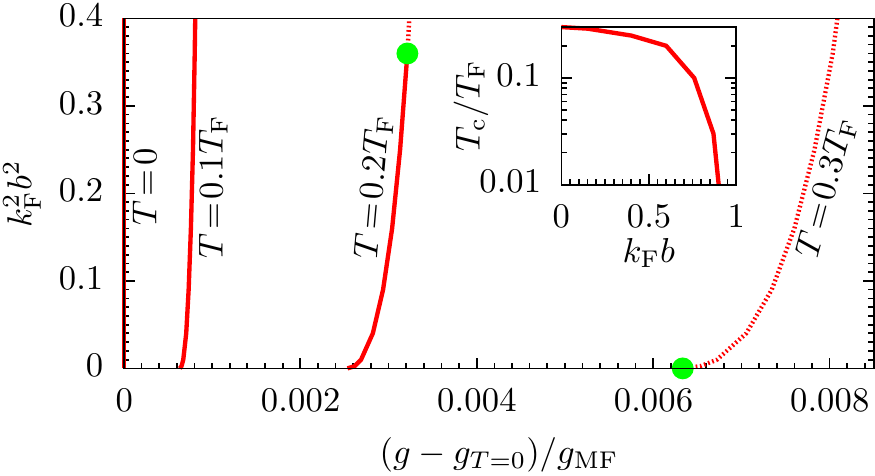}
 \caption{(Color online) (a) Phase boundary for the emergence of
   ferromagnetism with varying interaction strength ($g/g_{\text{MF}}$,
   primary x-axis) and screening length ($k_{\text{F}}^2b^2$, primary
   y-axis) at zero temperature. The secondary x-axis and y-axis show the
   parameters relevant for cold atom gases. The solid line corresponds to a
   first order transition, and the dashed line a second order
   transition. The red line shows the boundary for the screened Coulomb
   interaction, and the blue line for the square potential. The magenta
   solid line shows the analytical result for the first order transition
   with a screened Coulomb interaction.  (b) The magnetization at the first
   order transition as a function of Coulomb screening length calculated
   with QMC (red) and analytically (blue). The green line shows the minimum
   magnetization of a first order transition that can be resolved within
   QMC.  (c) The shift in the phase boundary interaction strength,
   $g-g_{T=0}$ from the screened Coulomb interaction zero temperature
   analytical phase boundary shown in ((a), magenta solid line) at finite
   temperatures $T\in\{0,0.1,0.2,0.3\}T_{\text{F}}$ calculated using the
   analytical formalism. The red solid lines denote a first order transition
   and the dotted red line a second order transition, and the green dots the
   tricritical points. The inset shows the tricritical point temperature
   with screening length.}
 \label{fig:EffectiveRange}
\end{figure}

With both the Quantum Monte Carlo and analytical formalism in place we are
well positioned to study the emergent phase diagram. We first focus on the
screened Coulomb potential relevant for the solid state, before looking at
the cold atom gas.

\subsection{Solid state}

We first study the screened Coulomb interaction relevant for the solid
state. With a short ranged interaction $k_{\text{F}}b\ll1$, the Fourier
transform is momentum independent so we recover the contact interaction
limit. We performed DMC calculations to determine the ground state energy
for several different values of magnetization and interaction strength,
plotting the energy bands in \figref{fig:EnergyDiagram}(a). The minimum in
each band reveals the magnetization at that interaction strength. With
rising interaction strength the magnetization jumps from zero to
$M\approx0.45$, characteristic of a first order transition. This gives the
magnetization with interaction strength curve in
\figref{fig:EnergyDiagram}(c) that demonstrates not only the first order
transition, but how the magnetization subsequently grows into the fully
polarized state. The prediction of a magnetic transition at
$g/g_{\text{MF}}=0.57$ is in good agreement with the \eqnref{eqn:FreeEnergy}
prediction of a first order transition at $g/g_{\text{MF}}=0.51$, that was
also found in a previous analytical study~\cite{Conduit2D}.

We next repeat the procedure with a larger screening length
$k_{\text{F}}^2b^2=0.2$. \figref{fig:EnergyDiagram}(b,d) shows that the
magnetization grows smoothly with interaction strength, demonstrating a
second order transition within the magnetization resolution of our QMC
calculations. We summarize our results for several screening lengths in the
phase diagram in \figref{fig:EffectiveRange}(a). This demonstrates how upon
increasing the screening range the critical interaction strength increases
and the transition reverts from first to second order. This trend is in good
agreement with the analytical predictions from \eqnref{eqn:FreeEnergy},
though the analytical formalism predicts that the transition remains first
order due to the presence of a logarithmic divergence in the free energy
$F\sim|M|^3\log(T)$~\cite{Conduit2D,Belitz05,Maslov06,Efremov08}.

In order to probe the question of the order of the transition more closely,
in \figref{fig:EffectiveRange}(b) we show the magnetization formed at the
first order transition as a function of the screening length
$k_{\text{F}}^2b^2$, predicted by both QMC and from the analytical
formalism. The analytical formalism predicts a magnetization that falls
rapidly with screening length because the screening reducing the fluctuation
correction term that was responsible for driving the first order transition
at a lower interaction strength, but the transition remains first order due
to the logarithmic divergence in the free energy. The magnetization
following a first order transition with the contact interaction is predicted
to be smaller by QMC, and drops to zero at $k_{\text{F}}^2b^2=0.2$, at which
point the transition becomes second order. However, since the QMC method is
able to sample the magnetization only at discrete values,
$M\in\{0,0.07,0.19,0.28,...\}$, the minimum magnetization that can be formed
in a first order transition is $M_{\text{min}}=0.19$. Due to the rapid decay
in the magnetization formed following the first order transition with
screening length, we cannot resolve whether QMC predicts a first or second
order transition at $k_{\text{F}}^2b^2>0.1$, however the predicted crossing
of the boundary $M_{\text{min}}=0.19$ predicted at
$k_{\text{F}}^2b^2\approx0.1$ by QMC and $k_{\text{F}}^2b^2\approx0.18$ from
analytics is consistent.

The DMC results have good qualitative agreement with our analytical
formalism at $T=0$. This gives us confidence to use the analytics to study
the phase behavior at finite temperature. \figref{fig:PhaseDiagram} and
\figref{fig:EffectiveRange}(c) show that the interaction strength of the
transition increases with temperature as the fluctuation correction term is
depressed~\cite{Conduit2D}. The trend remains the same with increasing
screening length. However, with increasing temperature the transition
reverts from first to second order, and the inset of
\figref{fig:EffectiveRange}(c) reveals that the tricritical point
temperature reduces markedly with increased screening length.

\subsection{Cold atom gas}

We now study the possible emergence of ferromagnetic correlations in an
ultracold atomic gas. The electrons are simulated by a two-component gas of
fermionic atoms, that are mapped onto pseudo up and down-spin species. The
interactions between the atoms are controlled using a Feshbach
resonance. This gives experimentalists access to not only a range of
different interaction strengths but also the effective range of the
interparticle potential~\cite{Kohstall12}, that can be either positive or
negative depending on both the elements and the Feshbach resonance used. In
a cold atom gas the parameter used to characterize the interaction strength
is the two-dimensional scattering length $a_{2\text{D}}$ that can be related
to the interaction strength used within the Coulomb interaction by
$\ln(k_{\text{F}}a_{2\text{D}})=g_{\text{MF}}/2\pi g$, and the effective
range is linked to the Coulomb interaction screening length by
$r_{\text{e}}=-b^2/a$. However, this gives us access to only negative
effective ranges $r_{\text{e}}\le0$.  To study a positive effective range we
turn to the square potential $U\theta(R-r)$ to model the interactions. Not
only will this give allow us to study positive effective range, but it will
also serve as a useful point of comparison to the results for the screened
Coulomb interaction.  The square potential has a two-dimensional scattering
length $a_{2\text{D}}$ given by
$\ln(k_{\text{F}}a_{2\text{D}})=(g_{\text{MF}}/8\pi^2R)/
[1-\tanh(\chi)/\chi]$ with $\chi=R\sqrt{U}$, and effective range
\begin{align}
 r_{\!\text{e}}\!=\!\frac{R}{6}
 \!\left(\!9\!+\!\!
 \left[\!1\!-\!\frac{\tanh\chi}{\chi}\right]^{\!-\!2}
 \!\!\!\!\!\!-\!3\left[\frac{1}{\chi}\!+\!4\right]\!\!\!
  \left[1\!-\!\frac{\tanh\chi}{\chi}\right]^{\!-\!1}
 \!\right)\!\!\punc{.}
\end{align}

We now use DMC calculations to determine the ground state in the presence of
the square well potential. This allows us to augment the phase diagram in
\figref{fig:EffectiveRange}. The critical interaction strength and the
emergence of the second order transition are similar for both the square
well potential and our previous study of the screened Coulomb interaction,
though the square potential can be extended to effectively negative
$k_{\text{F}}^2b^2$. This verifies the form of our phase diagram, and
confirms that irrespective of the inter-particle potential our main result
of the transition changing from first to second order with rising screening
length within the resolution of our simulations, accompanied by a rise in
the critical interaction strength is robust.

The persistence of the ferromagnetic transition from negative into positive
effective ranges $k_{\text{F}}r_{\text{e}}$ makes the two-dimensional gas a
tantalizing target in the search for ferromagnetism in a cold atom
gas. Recent experiments have shown that atomic gases with a large negative
effective range~\cite{Kohstall12} have greatly suppressed two and three-body
losses. Our predictions of the changing magnetization in
\figref{fig:EnergyDiagram} and \figref{fig:EffectiveRange}(b) could be
measured through phase contrast imaging that would reveal the growing
polarization in the emerging domains~\cite{Conduit09ii}. Our accurate
determination of the phase boundary through DMC calculations, complemented
by analytical predictions, should guide future experiments.

\section{Discussion}

In this paper we have outlined Quantum Monte Carlo calculations on the
two-dimensional itinerant ferromagnet. The calculations show that the system
undergoes a paramagnet to ferromagnet phase transition. The transition is
first order for interactions with a short effective range, and within our
magnetization resolution second order for interactions characterized by
longer effective ranges. We have reconciled this with the damping effect
that long range interactions have on the quantum fluctuations. We have
compared the DMC results to an analytical formalism and found good
agreement, allowing us to then use the analytic formalism to extend the
phase diagram to finite temperatures. Higher temperatures suppress the first
order behavior leading to a tricritical point at $T=0.3T_{\text{F}}$, whose
temperature reduces with increasing screening length.

With typical solid state experimental systems having a screening length
$0.5\le k_{\text{F}}b\le 1$~\cite{Mazin97,Santil01}, the first order
transition should still be visible but with the fluctuation correction term
tempered, and thus the associated tricritical temperature will be reduced.
This could reduce the tricritical point seen in theory $0.3T_{\text{F}}$ to
the $0.02T_{\text{F}}$ seen in experiments on both two (and three)
dimensional
materials~\cite{Ikeda97,Mazin97,Perry04,Kikugawa04,Nakamura02,Saxena00,Huxley01,Watanabe02}
and increase the visibility of exotic low temperature phases that are
promoted by quantum criticality such as the spin spiral and p-wave
superconductor.

\acknowledgments {The author thanks Stefan Baur, Nigel Cooper, Andrew Green,
  Chris Pedder, and Curt von Keyserlingk for useful discussions, and
  acknowledges the financial support of Gonville \& Caius College.}


\begin{thebibliography}{99}

\bibitem{Conduit2D}
G.J.~Conduit, Phys.~Rev.~A {\bf 82}, 043604 (2010).

\bibitem{Hertz1976}
J. Hertz, Physical Review B {\bf 14},  1165  (1976).

\bibitem{Efremov08}
D.V.~Efremov, J.J.~Betouras, and A.~Chubukov, Phys. Rev. B {\bf 77},
220401(R) (2008).

\bibitem{Belitz05}
D.~Belitz, T.~Kirkpatrick, and T.~Vojta, Rev. Mod. Phys. {\bf 77}, 579 (2005).

\bibitem{Maslov06}
D.L.~Maslov, A.V.~Chubukov, and R.~Saha,
Phys. Rev. B {\bf 74}, 220402(R) (2006).

\bibitem{Conduit09}
G.J.~Conduit, A.G.~Green, and B.D.~Simons,
  Phys. Rev. Lett. {\bf 103}, 207201 (2009).

\bibitem{Conduit08}
G.J.~Conduit and B.D.~Simons, Phys. Rev. A \textbf{79},
053606 (2009).

\bibitem{Karahasanovic12}
U.~Karahasanovic, F.~Kr\"uger, and A.G. Green, Phys. Rev. B {\bf 85}, 165111 (2012).

\bibitem{Ikeda97}
S.~Ikeda, Y.~Maeno, and T.~Fujita, Phys. Rev. B {\bf 57}, 978 (1998).

\bibitem{Perry04}
R.S.~Perry, K.~Kitagawa, S.A.~Grigera, R.A.~Borzi, A.P.~Mackenzie,
K.~Ishida, and Y.~Maeno, Phys. Rev. Lett. {\bf 92}, 166602 (2004).

\bibitem{Mazin97}
I.I.~Mazin and D.J.~Singh, Phys. Rev. Lett. {\bf 79}, 733 (1997).

\bibitem{Kikugawa04}
N.~Kikugawa, C.~Bergemann, A.P.~Mackenzie, and Y.~Maeno, Phys. Rev. B {\bf 70}, 134520 (2004).

\bibitem{Nakamura02}
F.~Nakamura, T.~Goko, M.~Ito \emph{et al}., Phys. Rev. B {\bf 65}, 220402 (2002).

\bibitem{Huxley01}
A.~Huxley \emph{et al.}, Phys. Rev. B {\bf 63}, 144519 (2001).

\bibitem{Watanabe02}
S.~Watanabe and K.~Miyake, J. Phys. Chem. Solids {\bf 63}, 1465 (2002).

\bibitem{Saxena00}
S.S. Saxena, P. Agarwal, K. Ahilan, F.M. Grosche,
R.K.W. Haselwimmer, M.J. Steiner, E. Pugh, I.R.
Walker, S.R. Julian, P. Monthoux, 
G.G. Lonzarich, A. Huxley, I. Sheikin, D. Braithwaite and J. Flouquet,
Nature (London) {\bf 406}, 587 (2000).

\bibitem{Jo09}
G.-B.~Jo \emph{et al.}, Science \textbf{325}, 1521 (2009).

\bibitem{Kohstall12}
C.~Kohstall, M.~Zaccanti, M.~Jag, A.~Trenkwalder, P.~Massignan, G.M.~Bruun, F.~Schreck, and R.~Grimm,
Nature {\bf 485}, 615 (2012).

\bibitem{Schirotzek09}
A.~Schirotzek, C.-H.~Wu, A.~Sommer, and M.W.~Zwierlein,
Phys. Rev. Lett. {\bf 102}, 230402 (2009).

\bibitem{Duine05}
R.A. Duine and A.H. MacDonald, Phys. Rev. Lett.
\textbf{95}, 230403 (2005).

\bibitem{LeBlanc09}
L.J.~LeBlanc, J.H.~Thywissen, A.A.~Burkov and A.~Paramekanti, Phys. Rev. A \textbf{80},
013607 (2009).

\bibitem{Conduit09ii}
G.J.~Conduit and B.D.~Simons, Phys. Rev. Lett. 103, 200403 (2009).

\bibitem{Pekker11}
D.~Pekker \emph{et al.}, Phys.~Rev.~Lett. {\bf 106}, 050402 (2011).

\bibitem{Sanner12}
C.~Sanner, E.J.~Su, W.~Huang, A.~Keshet, J.~Gillen, and W.~Ketterle,
Phys. Rev. Lett. {\bf 108}, 240404 (2012).

\bibitem{vonKeyserlingk11}
C.W.~von~Keyserlingk and G.J.~Conduit,
Phys. Rev. A {\bf 83}, 053625 (2011).

\bibitem{Conduit10}
G.J.~Conduit and E.~Altman, Phys.~Rev.~A {\bf 82}, 043603 (2010).

\bibitem{Baur12}
S.K.~Baur and N.R.~Cooper, arXiv:1208.6540.

\bibitem{Bugnion13i}
P.O.~Bugnion and G.J.~Conduit, arXiv:1304.3299 (2013).

\bibitem{Bugnion13ii}
P.O.~Bugnion and G.J.~Conduit, arXiv:1304.3323 (2013).

\bibitem{Thankappan93}
V.K.~Thankappan, Quantum Mechanics, New Age International, (1993).

\bibitem{Needs10}
R.J.~Needs, M.D.~Towler, N.D.~Drummond, and P.~L\'opez~R\'ios,
J.~Phys.:~Condensed~Matter {\bf 22}, 023201 (2010). 

\bibitem{Ceperley80}
D.M.~Ceperley and B.J.~Alder, Phys. Rev. Lett. {\bf 45}, 566 (1980).

\bibitem{Ortiz99}
G.~Ortiz, M.~Harris, and P.~Ballone, {\bf 82}, 5317 (1999).

\bibitem{Zong02}
F.H.~Zong, C.~Lin, and D.M.~Ceperley, Phys. Rev. E {\bf 66}, 036703 (2002).

\bibitem{Pilati10}
S.~Pilati, G.~Bertaina, S.~Giorgini, and M.~Troyer, Phys. Rev. Lett. {\bf 105}, 030405 (2010).

\bibitem{Chang11}
S.-Y.~Chang, M.~Randeria, N.~Trivedi, PNAS, {\bf 108}, 51 (2011). 

\bibitem{Drummond11}
N.D.~Drummond, N.R.~Cooper, R.J.~Needs, and G.V.~Shlyapnikov, Phys. Rev. B
{\bf 83}, 195429 (2011).

\bibitem{Needs08}
R. Needs, M. Towler, N. Drummond and P. L\'opez  R\'ios,
CASINO version 2.3 User Manual, Cambridge University (2008);
W.M.C. Foulkes, L. Mitas, R.J. Needs and G. Rajagopal,
Rev. Mod. Phys. {\bf 73}, 33 (2001).

\bibitem{Roothaan60}
C.C.J.~Roothaan, Rev. Mod. Phys. {\bf 32}, 179 (1960).

\bibitem{Drummond04}
N.D. Drummond, M.D. Towler, and R.J. Needs, Phys. Rev. B
{\bf 70}, 235119 (2004).

\bibitem{LopezRios06}
P.~L\'opez R\'ios, A.~Ma, N.D.~Drummond, M.D.~Towler, and R.J.~Needs,
Phys. Rev. E {\bf 74}, 066701 (2006).

\bibitem{Umrigar07}
C.J.~Umrigar, J.~Toulouse, C.~Filippi, S.~Sorella, and R.G.~Hennig, 
Phys. Rev. Lett. {\bf 98}, 110201 (2007).

\bibitem{Kalos62}
M.H.~Kalos, Phys. Rev. {\bf 128}, 1791 (1962).

\bibitem{Kalos67}
M.H.~Kalos, J. Comput. Phys. {\bf 2}, 257 (1967).

\bibitem{Kalos74}
M.H.~Kalos, D.~Levesque, and L.~Verlet, Phys. Rev. A {\bf 9}, 257 (1974).

\bibitem{Ceperley81}
D.M.~Ceperley, M.H.~Kalos, and J.L.~Lebowitz, Macromolecules, {\bf 14}, 1472 (1981).

\bibitem{Reynolds82}
P.J.~Reynolds, D.M.~Ceperley, B.J.~Alder, and W.A.~Lester, J. Chem. Phys. {\bf 77}, 5593 (1982).

\bibitem{Ceperley84}
D.M.~Ceperley and B.J.~Alder, J. Chem. Phys. {\bf 81}, 5833 (1984).

\bibitem{Ceperley86}
D.M.~Ceperley and M.H.~Kalos, {\it ``Monte Carlo Methods in Statistical Physics, 2nd ed.''}, edited by K.~Binder, Springer, Berlin (1986).

\bibitem{Schmidt87}
K.E.~Schmidt and M.H.~Kalos, {\it ``Applications of Monte Carlo Methods in Statistical Physics, 2nd ed.''}, edited by K.~Binder, Springer, Berlin (1987).

\bibitem{Foulkes01}
W.M.C.~Foulkes, L.~Mitas, R.J.~Needs, and G.~Rajagopal, Rev. of Mod. Phys, {\bf 73}, 33 (2001).

\bibitem{Lee11}
R.M.~Lee, G.J.~Conduit, N.~Nemec, P.~L\'opez~R\'ios, and N.D.~Drummond,
Phys. Rev. E {\bf 83}, 066706 (2011).

\bibitem{Drummond08}
N.D.~Drummond, R.J.~Needs, A.~Sorouri, and W.M.C.~Foulkes,
Phys. Rev. B {\bf 78}, 125106 (2008).

\bibitem{Baldereschi73}
A.~Baldereschi, Phys. Rev. B {\bf 7}, 5212 (1973).

\bibitem{Chiesa06}
S.~Chiesa, D.M.~Ceperley, R.M.~Martin, and M.~Holzmann, Phys. Rev. Lett. {\bf 97}, 076404 (2006).

\bibitem{Holzmann09}
M.~Holzmann, B.~Bernu, V.~Olevano, R.M.~Martin, and D.M.~Ceperley, Phys. Rev. B {\bf 79}, 041308(R) (2009).

\bibitem{Zong03}
F.H.~Zong, D.M.~Ceperley, S.~Moroni, and S.~Fantoni, Mol. Phys. {\bf 101}, 1705 (2003).

\bibitem{Holzmann06}
M.~Holzmann, B.~Bernu, and D.M.~Ceperley, Phys. Rev. B {\bf 74}, 104510 (2006).

\bibitem{Drummond09}
N.D.~Drummond, and R.J.~Needs, Phys. Rev. B {\bf 79}, 085414 (2009).

\bibitem{Carleo11}
G.~Carleo, S.~Moroni, F.~Becca, and S.~Baroni, Phys. Rev. B {\bf 83}, 060411(R) (2011).

\bibitem{Hubbard79}
J.~Hubbard, Phys. Rev. B {\bf 19}, 2626 (1979).

\bibitem{Prange79}
R.E.~Prange and V.~Korenman, Phys. Rev. B {\bf 19}, 4691 (1979).

\bibitem{vonKeyserlingk13}
C.W.~von~Keyserlingk and G.J.~Conduit,
arXiv:1301.6036, submitted to Phys.~Rev.~B (2013).

\bibitem{Santil01}
G.~Santi, S.B.~Dugdale and T.~Jarlborg, 
Phys. Rev. Lett. {\bf 87}, 247004 (2001).

\end{thebibliography}
\end{document}